\documentclass[12pt]{article}
\usepackage{latexsym}
\usepackage{geometry} 
\usepackage{pict2e}

\def\bs{\begin{subequations}}
\def\es{\end{subequations}}

\catcode`\@=11

\newtoks\@stequation
\def\subequations{\refstepcounter{equation}
  \edef\@savedequation{\the\c@equation}%
  \@stequation=\expandafter{\theequation}
  \edef\@savedtheequation{\the\@stequation}
  \edef\oldtheequation{\theequation}%
  \setcounter{equation}{0}%
  \def\theequation{\oldtheequation\alph{equation}}}

\def\endsubequations{\setcounter{equation}{\@savedequation}%
  \@stequation=\expandafter{\@savedtheequation}%
  \edef\theequation{\the\@stequation}\global\@ignoretrue}

\makeatletter
        \renewcommand{\theequation}{\thesection.\arabic{equation}}%
        \@addtoreset{equation}{section}%
\makeatother

\usepackage{hyperref}

\begin{document}
\begin{titlepage}
\begin{center}
{\bf {A Consistent Theory of Tachyons with Interesting Physics for Neutrinos}}
\vskip 2cm
Charles Schwartz  \\
Department of Physics, University of California, Berkeley, CA. \\
 schwartz@physics.berkeley.edu,    May 26, 2022 \\

\vskip 1cm
Revised manuscript Submitted to the MDPI journal Symmetry, by invitation \\
for the Special Issue, Tachyons and Fundamental Symmetries
  
 \end{center}
 
\vfill

\begin{abstract}
Working strictly within the physics theories of Special and General Relativity, I have produced a series of studies developing a consistent mathematical description of tachyons, using both classical and quantum frameworks for particles and fields. The most important choices throughout this work concern the question of which habits from the study of ordinary particles (those that are restricted to velocities less than that of light) should be kept and which should be changed. The first part of this paper notes an alternative set of theories wherein that question is answered differently from the choices of this author; and the results of that are severe in terms of physical symmetry. Following that is a broad summary of what has been accomplished in this work: this starts with the recognition that low energy tachyons will create large gravitational fields through the space-components of the energy-momentum tensor and leads to studying properties of the cosmic neutrino background.  Lastly there is a discourse on the various arguments that have been given against the very possibility for tachyons to exist.

\end{abstract}

\vfill 
\end{titlepage}

\section{Introduction}

This is a special issue of the journal SYMMETRY, devoted to the topic of TACHYONS.  I am a theoretical physicist (a rather old one) and I have been invited to write a paper for this volume. So, I feel obliged to start out addressing the question, "What is a Tachyon?"  \footnote{At an early stage of this project I asked the editors if they would start the volume off with their own discussion of this question, offering a suggestion on how that might be done. Their response was rather empty; and so I am undertaking this task myself.}

Commonly stated, "Tachyons" are particles that travel at speeds greater than the speed of light. 

Commonly believed, Einstein's Theory of Relativity said that particles could never travel at speeds greater than the speed of light.

Physics is a human enterprise composed of people who do actual measurements of physical systems (simple or complex) who are called "experimentalists"; and also people, called "theorists",  who draw mathematical frameworks that they can attribute to experimental observations.

The overall consensus of physicists is that there has been no credible observation of anything traveling faster than the speed of light (in vacuum). Nevertheless, some have been enticed to consider that such things may exist. This is not a new field of speculation, it is not a popular field of speculation, but it is a lively topic for some few theorists.

Let me start by clarifying "what Einstein said." Without searching back into all the writings of that great physicist, I will state what I believe to be the strong consensus among today's physicists. According to Einstein's Special Theory of Relativity, any object (huge as a sun or tiny as an atom) that can be found at rest (by us observers) can also be seen as moving with some velocity v but there is a limit: v is always less than c, where c is the speed of light. Light itself always travels at exactly that speed c, never more never less.. (We are looking at motion of objects in otherwise empty space, with no external interactions, and not any gravity.) The mathematical formulation of the Special Theory of Relativity allows the possibility that there might be some other type of particles that would always travel at speeds greater than c; and such things are named "tachyons".

To get some more precise definition of the idea "tachyon", I would start by appealing to the broad concept of SYMMETRY, which is the title of this journal. Among the profession of particle physicists, I would assert that the best definition of the concept "particle" is associated with the mathematical prospect of finding the "irreducible representations of the Poincar\'e group." 

What is that? Wikipedia \cite{Wi} says, ``The Poincar\'e group, named after Henri Poincar\'e (1906),was first defined by Hermann Minkowski (1908) as the group of Minkowski spacetime isometries. It is a ten-dimensional non-abelian Lie group that is of importance as a model in our understanding the most basic fundamentals of physics.''  That group is a mathematical formalism that expresses a set of physical symmetries. We have, following Einstein, a world of four dimensions (three dimensions of space plus time) with the following symmetries for the vacuum (assumed empty space in which we may see free particles of all kinds and their interactions). Symmetry of translation,  mathematically generated by the four operators $P^\mu = (P^0, P^1, P^2, P^3) = (E, \textbf{p})$ and called "momentum"; symmetry of rotation, mathematically generated by the three operators $J_i $ and called angular momentum; symmetry of transition to a moving reference frame, mathematically generated by the three operators $K_i$ and called boosts. Those six operators, for rotations and boosts, comprise the Lorentz group, a subgroup of the Poincar\'e group.

The mathematical analysis of this Poincar\'e group leads first to the recognition that the four momentum operators $P^\mu$ all commute with one another and this leads to constructing plane waves as the proper basis for any free particle. The angular momentum operators and the boost operators may mix the components of $P^\mu$ but there is one simple combination that is unchanged (invariant) under those operations: it is written as $P^\mu P_\mu$

The idea of "irreducible representations" means that any physical system, viewed as moving freely in otherwise empty space-time, must be characterized by a constant value of $P^\mu P_\mu = E^2 - p^2$. That constant is a number associated with this particular physical system (and the word "mass" comes in here); and we see three distinct categories. These may be characterized by the range of velocities, where we recognize for any free particle  $v = p/E$. (Our units have the velocity of light c = 1.)  
\begin{eqnarray}
P^\mu P_\mu >0, ordinary\; particles\; with\;v < 1, \\
P^\mu P_\mu = 0, particles\; of\;light \;(photons)\; with \;v = 1, \\
P^\mu P_\mu<0,  some\;other \;type\;of\;particle \; with \;v > 1.
\end{eqnarray}
The word "tachyon" is commonly attached to this third category of particles. That is our first step in defining "tachyon". However, when we look at how various theoretical studies have proceeded, it becomes clear that there are more than one set of mathematical rules being taken to define what is a "tachyon." This distinction is particularly vivid in working with a quantum theory of tachyon fields.  I would offer two classifications of such theories.

{\bf {Def. A}} Tachyon is a particle (or field) that always travels (propagates) outside of the light cone ($v > 1$).

{\bf {Def. B}} Tachyon is a particle that may be derived from a wave equation (like Klein-Gordon or Dirac) where the usual mass parameter is replaced by an imaginary number.

These two definitions share some domain of overlap; but they can lead to very different physical theories. My own work is within the definition {\bf A}. There is a large body of work by other authors who start with the definition {\bf B}, follow traditional lines of going from classical physics to quantum physics, and then find it necessary to break the symmetry of the Lorentz group. My own work, retains strict adherence to the full Lorentz and Poincar\'e groups; thus there is substantial conflict between these two arenas of tachyon theory. If you want to see some of the clash between these two sets of theories, I offer this transcript of letters related to the peer review of my most recent paper.  \url{https://www.ocf.berkeley.edu/~schwrtz/TachyonDebate.pdf}

To illustrate the pivotal question of how one should go from accepted theory of ordinary particles to a theory of tachyons, here is a simple example which is basic to any physicist: How do you count the number of particles? For ordinary particles, especially in their rest frame, the answer is: Count the number of  particles $N$  in this particular Volume V at time t. 

Mathematically, and working within the Special Theory of Relativity, we assert that there is vector current density $j^\mu (x) = j^\mu(\textbf{x},t)$ that is locally conserved: 
\begin{equation}
\partial_\mu j ^\mu (x) = \partial /\partial t\;j^0 +\sum_{i=1}^3\partial/ \partial x^i\; j^i = 0. 
\end{equation}
Then we integrate this equation over some chosen volume ${\cal{V}}$ in four-dimensional space-time. 
For ordinary matter (that which travels at speeds less than c) we find it convenient to choose that volume as bounded between the two times $t_1 < t_2$, and then also bounded by a space-like surface S. Using Gauss's Theorem, the resulting identity is
\begin{equation}
\int d^3x\;  j^0(\textbf{x}, t_2) - \int d^3x\; j^0 (\textbf{x}, t_1) = 0.
\end{equation}
Here is the definition of the number of particles $N(t)$ and it is independent of time. In getting this result we have dropped any terms that sit on that other surface S. This is because we assume that the particles (or the field) is confined to some finite domain of 3-space at time $t_1$; and then, allowed to propagate only at speeds limited by the speed of light, it will be confined to a finite domain of 3-space at all times up to $t_2$. We locate that other surface S far enough out so that there is no matter there at any time of interest. 

Now we move on to consider tachyons. These are particles whose velocities are unlimited; so the above construction is false. We still have a locally conserved vector current density; and need to find a different arrangement for the integral in 4-space.\cite{CS7} It turns out that the proper way to count tachyons is to ask: How many particles flow through a fixed area in 3-space, integrating over time. For example,
\begin{equation}
N(z) = \int dt \int dx \int dy\; j^3 (x,y,z,t);
\end{equation}
and this will be independent of the value of z. 

I noted earlier that a major difference between my work on tachyons and that of numerous other authors involves how one goes from classical physics to quantum mechanics. The textbook answer, called "canonical quantization", is the path used by those other authors, following {\bf {Def. B}}. They find problems dealing with negative frequency solutions of the wave equations; and their choice is to abandon the symmetry of Lorentz transformations. The interested reader will undoubtedly find that line of work extensively presented in articles expected in this Special Issue of the journal Symmetry.\cite{ST}  In my own work (see reference \cite{CS2}) I discussed this matter in detail (it is analogous to the discussion above about N), I rejected "canonical quantization" and found an alternative route to a quantized field theory: simply relying upon  {\bf {Def. A}}. 

In response to some reviewers' comments, I shall briefly repeat, here, those arguments against using "canonical quantization" for tachyon field theory. The word "canonical" means following the structure of Hamiltonian dynamics in classical physics. The philosophical basis of that theory is to think that we can follow the evolution of any system as time evolves. That way to go from classical to quantum mechanics was first formulated for non-relativistic particle theory and then for relativistic field theory. We are familiar with the Hamiltonian H as the operator that generates time evolution of the physical system. As long as we limit our considerations to fields that propagate within (or on) the light-cone, we can have a well defined system contained in some finite region of space. (This is like the discussion, above, about the number of particles.) However, if we now consider tachyonic fields, this is no longer true. The mathematical question, similar to what I have detailed above, is how to go from a locally conserved energy-momentum tensor, $\partial_\mu T^{\mu \nu} = 0$, to a definition of the time-independent energy (H=E).

I have another way of distinguishing that other line of tachyon theory from my own work: It seems that the objective of those other authors is to take some equation that looks like it implies faster-than-light particles (e.g., start with {\bf {Def. B}}) and then manipulate the mathematics to get rid of troubles that they see and thereby restore the familiar behavior of slower-than-light particles, thus negating {\bf {Def. A}}. A recent paper in the journal Particles \cite{LN} is explicit about that mission. My  own mission, I  can state clearly, is to start with {\bf{Def. A}}, stay within the strict confines of traditional theories of Relativity in both Classical and Quantum Mechanics, and see where careful exploration may lead. This has been quite successful, as readers will see below. but much remains to be studied.

 Ultimately, it is experimental data that will determine the validity or irrelevance of any theoretical offerings. The experiment most currently watched is called KATRIN. This measures the energy spectrum of electrons from the beta-decay of Tritium and may set some limits on neutrino mass; and may possible distinguish among current theories. Otherwise, it is hard to see experiments that will directly say how fast low energy neutrinos move. 
 
 There are, however, other types of experimental observations that may be cited as justifying, or contradicting, certain theories about neutrinos as tachyons. My own work has found two such rewarding situations. In  my 2017 paper \cite{CS3} I reported that my theory of gravitation produced by a sea of tachyons could explain - qualitatively and quantitatively - the famous phenomenon known as Dark Energy. In my 2022 paper \cite{CS6} I came up with a model of higher spin tachyons that could explain - in rough quantitative agreement - the known mass hierarchy of the three types of neutrinos. My search for a detailed model of how tachyon neutrinos might provide an explanation for the phenomenon known as Dark Matter remains an open question.

What follows is a selection of my writings about tachyons. Section 2, which comes from a talk given at a physics conference in 2019,  is meant to briefly summarize my work, over the last decade, to construct a consistent mathematical theory for the idea that tachyons - faster then light particles - might exist in this universe. The physics here focuses on neutrinos and their role in Gravitation and  Cosmology. Section 3 is a polemical note entitled "Debunking the Anti-Tachyon Myths"; and it is addressed to a wide arena of physicists and others who believe that tachyons could not exist. 

\section{If Neutrinos are Tachyons, we may explain Dark Energy and Dark Matter }

What this section covers is a slightly edited version of a talk given at the Martin B. Halpern Memorial Conference, UC Berkeley, March 30, 2019. This is a quick survey of my previous research publications on a theory of tachyons. The several subsections are titled, OPERA, Classical Tachyons, Negative Energy States, Energy-Momentum Tensor I (DM), Energy-Momentum Tensor II (DE), Quantized Dirac Tachyon Field, Future Work. 
 
\subsection{OPERA}
Start with the OPERA experiment of 2011.
20 GeV neutrinos appeared to travel faster than light by about 1/40,000.

From the formula $E = mc^2/\sqrt{v^2/c^2 -1}$,
\begin{equation}
(v-c)(v+c)/c^2 = (mc^2/E)^2.
\end{equation}
So that would imply a tachyon neutrino with a mass of about 100 MeV. But we are pretty sure that neutrino mass is around 0.1 eV. So the OPERA result going away tells us nothing about possible neutrino tachyons of such a low mass. That is what we are considering.

For a comprehensive review of experimental data linked to the idea of tachyons, I recommend the article by Robert Ehrlich in this Special Issue. \cite{RE}

\subsection{Classical Tachyons}
Classical particles are described by a worldline $\xi^\mu(\tau)$ with a Lorentz invariant  form that leads us to note three categories:
\begin{eqnarray} 
\eta_{\mu \nu}\;\dot{\xi}^\mu\; \dot{\xi}^\nu = \epsilon \;\;\;\;\;\;\;\;\; \\
ordinary \; particles \; (v < c): &&\epsilon = +1 \nonumber\\
massless \; particles\; (v=c): &&\epsilon = 0 \nonumber\\
tachyons\; (v > c): &&\epsilon = -1 \nonumber
\end{eqnarray}

We usually define the 4-vector $p^\mu = m \dot{\xi}^\mu$. For tachyons this is a space-like vector and so one asks, What about negative energy states?

\subsection{Negative Energy States}
Let's look at a space-time diagram for a general interaction process. Figure 1 shows four particles involved in the reaction, $n \rightarrow p + e + \nu$, where I imagine that the neutrino is a tachyon.

\begin{picture}(100,200)
\thicklines 
\put(50,100){\circle*{10}}
\put(55,100){\line(1,1){60}}
\put(55,100){\line(1,-1){60}}
\put(45,100){\line(-1,1){60}}
\put(45,100){\line(-1,-1){60}}
\thinlines 
\put(50,105){\line(0,1){40}}
\put(50,95){\line(0,-1){40}}
\put(50,100){\line(-1,2){20}}
\put(55,100){\line(5,1){40}}
\put(50,150){\shortstack{p}}
\put(48,45){\shortstack{n}}
\put(25,145){\shortstack{e}}
\put(97,105){\shortstack{$\nu$}}
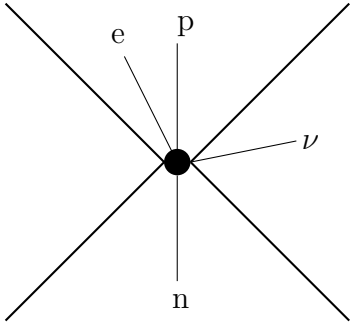
\put(0,20){\shortstack{Figure 1. Reaction with an outgoing tachyon.}}
\end{picture}

Now look at Figure 2. Is this a picture of the reaction $ n \rightarrow p + e + \nu$ with the neutrino carrying off negative energy; or is this a picture of the reaction $n + \nu \rightarrow p + e$ with positive energy for all participants?

\begin{picture}(100,200)
\thicklines 
\put(50,100){\circle*{10}}
\put(55,100){\line(1,1){60}}
\put(55,100){\line(1,-1){60}}
\put(45,100){\line(-1,1){60}}
\put(45,100){\line(-1,-1){60}}
\thinlines 
\put(50,105){\line(0,1){40}}
\put(50,95){\line(0,-1){40}}
\put(50,100){\line(-1,2){20}}
\put(55,100){\line(5,-1){40}}
\put(50,150){\shortstack{p}}
\put(48,45){\shortstack{n}}
\put(25,145){\shortstack{e}}
\put(97,90){\shortstack{$\nu$}}
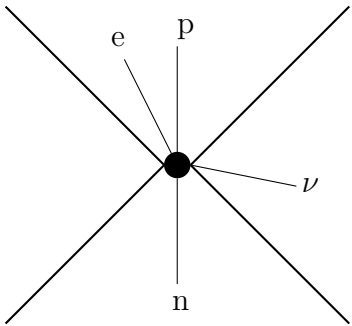
\put(0,20){\shortstack{Figure 2. Reaction with an incoming tachyon.}}
\end{picture}

This "problem" is the same as noting that the mass-shell equation,
\begin{equation}
p^\mu \; p_\mu = E^2 - p^2 = \pm m^2, \label{b1}
\end{equation}
gives us two separate hyperboloids for ordinary particles (plus sign) but a single hyperboloid for tachyons (minus sign). For ordinary particles, we manage to reinterpret the negative energy solutions as antiparticles and give them positive energy. For tachyons we need to see what it means when we look at a positive energy particle from a different Lorentz frame, where it appears to have negative energy.

Look again at the space-time diagrams above. A "positive energy" tachyon will have $dt/d\tau > 0$ and its trajectory will be seen moving upward - as in Figure 1 - and so we would call that an outward going particle if it sits above the interaction region in time; and we would call it an inward moving particle if the trajectory sits below the interaction region - as in Figure 2. But from another reference frame we may have $dt/d\tau < 0$ for what was formerly an outgoing particle and so this will now look like an incoming particle. 

The lesson is that \emph{the labels "in" and "out" are Lorentz invariant for ordinary particles but NOT for tachyons.}  Does this matter? No. The physical law which we call the conservation of total energy and momentum is written,
\begin{equation}
\sum_{out} p^\mu _j  - \sum_{in} p^\mu_i = 0.\label{b2}
\end{equation}
This is true in any Lorentz frame, even though the individual terms in this equation each transform. In going from Figure 1 to Figure 2 we need only move one $p^\mu$ from the "out" summation to the "in" summation.

\subsection{Energy-Momentum Tensor I (DM)}
Now we look at tachyons in General Relativity.

For ordinary particles we write the Energy-Momentum tensor as,
\begin{equation}
T^{\mu \nu} (x) = m\; \int d\tau\; \dot{\xi}^\mu\; \dot{\xi}^\nu \; \delta^4 (x -\xi(\tau)), 
\end{equation}
and, at first, I used this same formula for tachyons.

For a free particle we have the familiar representation,
\begin{equation}
\xi^\mu (\tau)= (\gamma \tau, \gamma \textbf{v}\;\tau), \;\;\;\;\; \gamma = 1/\sqrt{|1-v^2/c^2|}.\end{equation}

Thus, for a very low energy ordinary particle ($v << c$) the dominant term is 
\begin{equation}
T^{00} (x)= m  \delta^3(\textbf{x} - \textbf{v} t).
\end{equation}

and for a very low energy tachyon ($v >> c$) the dominant terms are
\begin{equation}
T^{i j} (x)= m \gamma v_i\; v_j\; \delta^3 (\textbf{x} - \textbf{v} t).
\end{equation}

These space-components of $T ^{\mu \nu}$ can be very large; and this leads to interesting cosmological modeling.  

In my 2011 paper I showed that this leads to attractive forces among co-linear flows of tachyons; and I predicted the possibility of such tachyon "ropes" becoming localized, say, within a galaxy, and creating strong local gravitational fields that could produce the observational effects now ascribed to Dark Matter.

Then, last year, I was led to revise that theory, as follows.
\subsection{Energy-Momentum Tensor II (DE)}

Let's derive that formula for tachyons' energy-momentum tensor from some general principle.  General Relativity may be constructed from an action principle with a Lagrangian density that looks like this.
\begin{eqnarray}
{\cal{L}} = \sum m\int d\tau\; \sqrt{\epsilon g_{\mu \nu}(x)\dot{\xi}^\mu\; \dot{\xi}^\nu }\delta^4 (x-\xi(\tau)) + \nonumber \\
\frac{ \sqrt{|det g|}}{8\pi G}\;R\;\;\;\;\;,
\end{eqnarray}
where we sum over all the particles and $R$ is the scalar form of the Riemann curvature tensor, which depends on the metric tensor $g_{\mu \nu}(x)$. Notice that I put an epsilon under the square root to make sure it would always be real, whether we have an ordinary particle or a tachyon.

GR textbooks show how, under variation of the metric $g_{\mu \nu}$, we get exactly Einstein's equation, with the energy momentum tensor, as we wrote it earlier, on the right hand side. Note, however, that the factor epsilon should be sitting there:
\begin{equation}
T^{\mu \nu} (x) = \epsilon m\; \int d\tau\; \dot{\xi}^\mu\; \dot{\xi}^\nu \; \delta^4 (x -\xi(\tau)).
\end{equation}

This new minus sign leads to the unconventional result of repulsive gravitational forces; and now we predict that tachyons, if they exist in the cosmos, would form a pervasive gas and contribute a \emph{negative pressure} when we consider the Robertson-Walker model for the universe.  The formula for this pressure is simply,
\begin{equation}
p = -m \gamma \frac{v^2}{3} \rho,
\end{equation}
where $\rho$ is the density of those tachyons, $v$ is their velocity, and $\gamma = 1/\sqrt{v^2/c^2 -1}$.

Using standard numbers for the density and energy of the Cosmic Neutrino Background, assuming that they could be tachyons with a mass around $ 0.1 \;eV/c^2$, gives a numerical value for \textbf{this negative pressure that is within a factor of two of explaining what is commonly called Dark Energy.}
\vskip 1cm 
Thus far, considering cosmic neutrinos as tachyons in General Relativity:

ACT I - this may explain Dark Matter

alternatively,

ACT II - this may explain Dark Energy

... waiting for Act III

\subsection{Quantized Dirac Tachyon Field}
Let's look at a quantized field theory for a Dirac tachyon.
\begin{equation}
\psi(x) = \int_{-\infty}^{\infty} d\omega \int d^2 \hat{k}\;k^2\; e^{i(\textbf{k}\cdot\textbf{x} - \omega t)} \sum_{h=\pm 1}\;b_h(\omega,\hat{k})\; v_h(\omega,\hat{k})
\end{equation}
where $k = \sqrt{\omega^2 + m^2}$, $\textbf{k} = k\;\hat{k}$, and $v_h$ is a Dirac spinor for a tachyon (put "im" instead of "m" in the Dirac wave equation), with helicity h. Note that the integral over frequencies extends over all real numbers. I do not separate positive and negative frequency solutions, since a Lorentz transformation of tachyon states readily mixes those two domains. (Please refer back to Section 1 of this paper where I discuss, and criticize, the work of other theorists who do make that separation, and break Lorentz invariance.) 

The operator algebra is chosen to be, 
\begin{equation}
[b_h(\omega, \hat{k}), b^\dagger _{h' }(\omega ', \hat{k}')]_+ = \delta(\omega - \omega') \delta^2(\hat{k} -\hat{k}')/k^2\; \delta_{h,h'};
\end{equation}
and this gives us the two-point anti-commutator for the fields,
\begin{equation}
[\psi(x),\psi^\dagger(x')]_+ = 0, \;\;\; if\;\;\; |t-t'| > |\textbf{x} - \textbf{x}'|.
\end{equation}
This is the strong condition of causality for tachyon fields: \emph{No signal can travel slower than the speed of light.}

Next, we look at one-particle states and then examine how to calculate the energy-momentum tensor.
What we note is that the helicity h serves to signify what we call the particle and what we call the anti-particle. (For ordinary particles this job was given to the sign of the frequency.) Working from the vacuum state $|0>$ we define:
\begin{equation}
b_{h=+1} (\omega, \hat{k}) |0> = 0, \;\;\; b^\dagger_{h=-1} (\omega, \hat{k}) |0> = 0.
\end{equation}
Then construct one-particle states,
\begin{eqnarray}
|\omega, \hat{k},h=+1> = \sqrt{k} \;b^\dagger_{h=+1}(\omega, \hat{k}) |0> , \\
|\omega, \hat{k},h=-1> = \sqrt{k} \;b_{h=-1}(\omega, \hat{k}) |0> .
\end{eqnarray}

The conserved energy-momentum tensor for the tachyon Dirac field is,
\begin{equation}
T^{\mu \nu} (x) = (i/4)\psi^\dagger (x)\gamma^0 \gamma^5 (\gamma^\mu \stackrel{\leftrightarrow}{\partial}^\nu + \gamma^\nu \stackrel{\leftrightarrow}{\partial}^\mu) \psi(x), 
\end{equation}
where $\stackrel{\leftrightarrow}{\partial}=\stackrel{\rightarrow}{\partial} - \stackrel{\leftarrow}{\partial}$.

I want to put this operator between one-particle states; but first there are two important steps. First we embrace this operator with the notation $: ... :$ which means "normal order", to guarantee that this will have zero expectation value in the vacuum state.

The second step is to acknowledge that we need a specific indefinite metric when taking matrix elements between one-particle states of spin one-half tachyons.  This comes from investigating the "Little Group" $O(2,1)$ appropriate for tachyons in building a unitary representation of the Lorentz group. This indefinite metric $H$ is simply the helicity operator.

The result of this calculation is,
\begin{equation}
<\omega, \hat{k},h|H :T^{\mu \nu}: |\omega, \hat{k},h> = h\; k^\mu\;k^\nu\;.
\end{equation}
This says that particle and anti-particle contribute to the energy-momentum tensor with \emph{opposite} signs. \textbf{Thus we see the possibility to explain both Dark Matter and Dark Energy with the proposition that Cosmic Background Neutrinos are tachyons, with a mass in the neighborhood of $0.1\; eV/c^2$.}

\subsection{Future Work}
{\bf More Work that Needs to be Done}

\underline{Questions about previous results/claims}

* Should chirality, rather than helicity, define particle vs anti-particle for tachyons?

* Is $k^2 \;d\omega\;d^2\hat{k} $ the correct density of states formula for tachyons?

* Can we show that gravity works to localize tachyon streams?

\underline{New areas to be explored}

* Fitting of tachyon neutrinos into the Standard Model of particles.

* Mass mixing for tachyon neutrinos

* Revise Standard Cosmology Theory for tachyon neutrinos (many sub-topics)

* New experiments to detect low energy tachyon neutrinos

\vskip 1cm

\underline{Summary of Cosmology results - Scaling of Energy and Pressure Components}

Radiation:  $\rho\sim\;a^{-4}, \;\;\;\;\; p \sim\;a^{-4}$

Cold Matter:   $\rho\sim\;a^{-3}, \;\;\;\;\; p \sim\;0$

Cosmo Const:  $\rho\sim\;a^{0}, \;\;\;\;\; p \sim\;a^{0}$

Cold Tachyon Neutrinos:  $\rho\sim\;0, \;\;\;\;\; p \sim\;a^{-1.5}\;\;$
     or    $\;\; p \sim\;a^{-2}$
\vskip 1cm 
\underline{Something we can check now?}

If low energy tachyon neutrinos cause Dark Matter effects, those effects should be lessened in the past when their Temperature was near or above 0.1 eV. Is there data on Dark Matter effects as a function of z?
\vskip 1cm
For slow particles under Newton's Gravity, 
\begin{equation}
H = \sum_a \frac{1}{2} m_a v_a^2 - \sum_{a < b} \frac{Gm_a m_b}{ |\textbf{x}_a - \textbf{x}_b|}. \nonumber 
\end{equation}

And from Einstein's GR we now get,
\begin{eqnarray}
H = -\sum_a \omega_a  m_a \gamma_a  - \sum_{a , b} \frac{G(\omega_a m_a  \gamma_a )(\omega_b m_b \gamma_b)}{ |\textbf{x}_a - \textbf{x}_b|} Z_{ab} ,\;\; \nonumber\\
Z_{ab}=  2 - 4 \textbf{v}_a \cdot \textbf{v}_b + (v_a^2 + v_b^2) - (\epsilon_a \gamma_a^2 + \epsilon_b \gamma_b^2 + 1)\times \; \nonumber \\
\;\times  [(1-\textbf{v}_a \cdot \textbf{v}_b)^2 - \frac{1}{2}(1-v_a^2)(1-v_b^2)]. \;\;\nonumber
\end{eqnarray} 
For a static system, including tachyons.

\vskip 1cm
"For every complex problem there is a simple solution. And it's always wrong."

--H. L. Mencken

 \section {Debunking the Anti-Tachyon Myths}

Some people believe that the Earth is flat; and they can see this is true with their own eyes. Physicists are convinced otherwise; and they can cite abundant evidence on their side.

However, when it comes to Tachyons (faster-than-light particles), a great many physicists believe that they do not, and some will say that they can not, exist; and a number of reasons are recited in support that that prejudice.

In my several papers exploring mathematical frameworks for how Tachyons might fit into physical theory and experiments I have taken the trouble to lay out careful reasoning to debunk those prejudices.  Do I have to review all those arguments in every new paper I write? This note is posted to serve that purpose.

Firstly, all my work is done strictly within the established mathematical frameworks of Special and General Relativity. (Some other authors have violated those bounds.)

The argument most commonly heard against the existence of tachyons (anything that travels faster than light) is this: If tachyons existed, one could, in principle, send a signal into the past and this would lead to an unbearable logical paradox. (This so-called paradox is sometimes named the "antitelephone.")

This claim has been debunked some time ago \cite{ER}; and here I shall give my own critique.
\vskip 0.5cm
\textbf{The phrase "send a signal into the past" is utter nonsense. Said more professionally, this phrase is an oxymoron, meaning that it is self-contradictory in terms of the language used.}
\vskip 0.5cm
Let me show this by means of some diagrams that are basic tools in connecting the mathematical principles of the Special Theory of Relativity to the way we physicists talk.  The picture below is a space-time diagram that shows two events, labeled A and B, each with a worldline showing the trajectory of a particle (or a signal) ending at that event.  The picture at A is described by the word "send" and this means that the particle moves farther and farther away from the location of event A, as time moves forward. The picture at B is described by the word "receive" and this means that the particle moves closer and closer to the location of event B, as time moves forward. Thus the words "send" and "receive" each have within their definition (at least as sensible physicists speak) a specific sense of time evolution. The phrase "send a signal into the past" explicitly violates this definition of the word "send."

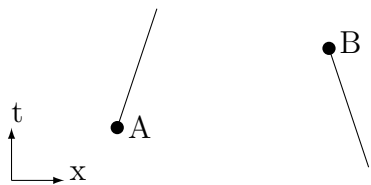
\begin{picture}(200,100)
\put(10,10){\vector(0,1){20}}
\put(10,32){t}
\put(10,10){\vector(1,0){20}}
\put(32,10){x}
\put(50,30) {\circle*{5}}
\put(54,26) {A}
\put(50,30){\line(1,3){15}}
\put(130,60){\circle*{5}}
\put(134,58){B}
\put(130,60){\line(1,-3){15}}
\end{picture}

Please note that I have not yet said anything about tachyons. This comes next.  

If the particle being sent or received is an ordinary particle (or a photon), then the trajectory will lie within (or on) the lightcone centered at the endpoint A or B. This property will remain the same when either process is viewed from a different Lorentz frame. Thus, for ordinary particles or light we can say that the assignment of the word "send" or "receive" is invariant under (orthochronous) Lorentz transformations. For tachyons, however, the particle trajectory lies outside the lightcone; and this means that a Lorentz transformation can change the appearance of the process. The experiment at A (sending the particle) may be seen by observers in a different Lorentz frame as a version of experiment B (receiving the particle).

Thus, one way to debunk the "antitelephone" paradox is to say that you have misused Special Relativity by  tieing together two observations that are actually made in different Lorentz frames and claiming this is physical reality. It is not. The error is to assume that the word "send" always has a Lorentz invariant meaning, when, in fact, that is true only for a select set of particles, excluding tachyons. ( There is some similarity between this discussion about "send" and "receive" and the discussion in Section 2.3 about "in states" and "out states".)

It may seem strange that, as one considers viewing the experiment at A from a sequence of Lorentz frames that move faster and faster away from the original frame, an experiment that was once termed "send" becomes instantaneously transformed into "receive". Next, I note a way to remove this annoyance.

In my 2011 JMP paper, \cite{CS1} Appendix A looks at a scenario of sending tachyon signals between earth and a distant rocket ship, alleging \emph{a causal paradox}. It is argued that an exchange of tachyon signals can lead to a response arriving before the original message was sent out. Simply replacing the point particle by a wave packet shows that, when one carries out the relevant Lorentz transformation, the distinction between sending (emitting) and receiving (absorbing) a tachyon can disappear in a continuous manner.

In my 2016 IJMPA paper,\cite{CS2} I state the appropriate \emph{principle of causality} for tachyons - no propagation slower than the speed of light; and this leads to a consistent mathematical formalism for quantizing such fields. This provides an alternative to the canonical formalism, which is wrong for tachyon fields.

In my 2018 IJMPA paper,\cite{CS4} Section 2 examines the role of tachyons engaged in a general multi-particle interaction. The common idea that \emph{negative energy states} imply physical instability of the system is debunked by recognizing that the naming of $in$ and $out$ states is not Lorentz invariant. The total energy and momentum are still conserved.

In my 2016 paper \cite{CS2} on quantizing tachyon fields, especially for the spin 1/2 (Dirac) case, I deal with the \emph{Little Group O(2,1)} by introducing an indefinite metric (the helicity) into the Fock space.

Then there are experiments, a number of which over the years have claimed to observe neutrinos as tachyons, and then been revised to the opposite conclusion. The 2011 \emph{OPERA experiment} looked at 20 GeV neutrinos and first reported that they travelled faster than light by 1 part in 40,000. That would imply a tachyon mass of about 100 MeV. But we know that neutrino mass is around 0.1 eV; and the excess velocity (v-c) goes as the square of the mass-to-energy ratio. That puts us 14 orders of magnitude below the original (wrong) observation.

Finally, there are theoretical efforts to derive the existence of known particles from some abstract field with complicated self-interactions. The simplest model is a scalar field with a potential that looks like W. If one expands around the central peak, then the resulting particles are found to be tachyons (negative mass-squared). But then one recognizes that those states are unstable; one should instead expand about the minima of W, where one gets ordinary particles.  I am not involved in that sort of theorizing. 

I start with the question: If tachyons do exist, how would we describe them within our customary mathematical frameworks? The starting point is the relativistically invariant form for any 4-vector (e.g., the energy-momentum of a particle): \newline $p^\mu p_\mu = constant$. That constant may be positive, zero, or negative.

 A recent criticism of tachyon-neutrinos is this. Take the Lagrangian density for an ordinary Dirac field,
 \begin{equation}
 \bar{\psi} [i\gamma^\mu \partial_\mu -m]\psi, \;\;\; \bar{\psi} = \psi^\dagger \gamma^0
 \end{equation}
 and replace $m$ by $im$. The result is non-Hermitian, so the theory derived from this will violate unitarity - conservation of total probability. But that is not how I do it. I insert $im$ in the Dirac differential equation, derive the equation for the adjoint field $\psi^\dagger$ from this, and then construct a Lagrangian density that will produce both of those equations of motion. The result is,
 \begin{equation}
 \bar{\psi} \gamma_5 [i\gamma^\mu \partial_\mu -im]\psi.
 \end{equation}
 With the $\gamma_5$ inserted this is all Hermitian. So there is no problem here.
 
 I have also noted earlier that the conserved current and energy-momentum tensors are,
 \begin{eqnarray}
 j^\mu = \bar{\psi} \gamma_5 \gamma^\mu \psi, \\
 T^{\mu \nu} = (im/4) \bar{\psi}\gamma_5 [\gamma^\mu \stackrel{\leftrightarrow}{\partial^\nu} + \gamma^\nu \stackrel{\leftrightarrow}{\partial^\mu} ]\psi.
 \end{eqnarray}

 While all these expressions are Hermitian and transform appropriately under proper orthochronous Lorentz transformations, they have the opposite behavior under parity (space inversion) compared to ordinary Dirac theory.  Is this a problem? Neutrinos are supposed to interact with other particles in a way that maximizes parity violation; so maybe this is a good sign! 
 
 Furthermore, given such forms, we would not want to couple this tachyon-Dirac field to the electromagnetic field. That is good news for our theory for two physical reasons: a charged tachyon would rapidly dissipate its energy via Cherenkov radiation; and neutrinos are uncharged particles.
 
 Let's continue this line of inquiry. Earlier I have said that it is the helicity of Dirac-tachyon states that designates the naming of particle vs anti-particle; and there has been some uncertainly about whether this is a Lorentz invariant rule.  Here is a short table of simple calculations of the scalar and the pseudoscalar forms for an ordinary Dirac wavefunction $\psi_o$ and for a tachyon Dirac wavefunction $\psi_t$, both sensibly normalized.
 \begin{equation}
\begin{array}{cll}
Scalar &  \bar{\psi}_o \psi_o = \omega/|\omega| = \pm 1 \;\;\;\;\;& \bar{\psi}_t \psi_t = 0 \\ 
 Pseudoscalar &  \bar{\psi}_o \gamma_5\psi_o = 0 & \bar{\psi}_t \gamma_5\psi_t = h = \pm 1 
\end{array}
\end{equation}
That speaks well for using helicity, $h$,  for tachyons as we use the sign of the frequency for ordinary Dirac particles.
  
  There is more we  might say. In the classical theory for tachyons I have written an additional factor $\zeta = \pm 1$ in front of the Lagrangian density.
  \begin{eqnarray}
  \zeta m \int d\tau \sqrt{\epsilon \dot{\xi}^\mu \dot{\xi}^\nu g_{\mu \nu}}\;\delta^4(x - \xi(\tau));
  \end{eqnarray}
  and I have said elsewhere that this sign factor should be the helicity of the particle state. That makes this expression a pseudoscalar, which aligns with the previous discussion.
  
  \vskip 1cm
  {\bf {Acknowledgment}}
  
  I am  grateful to Professor Robert Ehrlich who encouraged me to write this paper.

\end{document}